\documentclass[a4paper,twocolumn,11pt,accepted=2021-08-24]{quantumarticle}
\pdfoutput=1
\usepackage[utf8]{inputenc}
\usepackage[english]{babel}
\usepackage[T1]{fontenc}
\usepackage{amsmath}
\usepackage{hyperref}
\usepackage{lipsum}

\usepackage[numbers,sort&compress]{natbib}

\usepackage{qcircuit}
\usepackage[dvips]{graphicx}
\usepackage{amsmath,amssymb,amsthm,mathrsfs,amsfonts,dsfont}
\usepackage{subfigure, epsfig}
\usepackage{braket}
\usepackage{bm}
\usepackage{multirow}
\usepackage{enumerate}
\usepackage{color}

\newcommand{\qi}[1]{{#1}}

\newtheorem{theorem}{Theorem}
\newtheorem{lemma}{Lemma}
\newtheorem{corollary}{Corollary}

\begin{document}


\title{Exploiting anticommutation in Hamiltonian simulation}



\author{Qi Zhao}
\email{zhaoq@umd.edu}
\affiliation{Joint Center for Quantum Information and Computer Science, University of Maryland, College Park, Maryland 20742, USA}
\author{Xiao Yuan}
\email{xiaoyuan@pku.edu.cn}
\affiliation{Center on Frontiers of Computing Studies, Department of Computer Science, Peking University, Beijing 100871, China}
\affiliation{Stanford Institute for Theoretical Physics, Stanford University, Stanford California 94305, USA}
\begin{abstract}
Quantum computing can efficiently simulate Hamiltonian dynamics of many-body quantum physics, a task that is generally intractable with classical computers. The hardness lies at the ubiquitous anti-commutative relations of quantum operators, in corresponding with the notorious negative sign problem in classical simulation. Intuitively, Hamiltonians with more commutative terms are also easier to simulate on a quantum computer, and anti-commutative relations generally cause more errors, such as in the product formula method. 
Here, we theoretically explore the role of anti-commutative relation in Hamiltonian simulation. 
We find that, contrary to our intuition, anti-commutative relations could also reduce the hardness of Hamiltonian simulation. Specifically, Hamiltonians with mutually anti-commutative terms are easy to simulate, as what happens with ones consisting of mutually commutative terms. 
Such a property is further utilized to reduce the algorithmic error or the gate complexity in the truncated Taylor series quantum algorithm for general problems. 
Moreover, we propose two modified linear combinations of unitaries methods tailored for Hamiltonians with different degrees of anti-commutation. We numerically verify that the proposed methods exploiting anti-commutative relations could significantly improve the simulation accuracy of electronic Hamiltonians. 
Our work sheds light on the roles of commutative and anti-commutative relations in simulating quantum systems. 

\end{abstract}

\maketitle

\section{Introduction}

It is a notoriously hard problem to classically simulate an arbitrary large many-body physics systems. As proposed by Feynman in 1982~\cite{Feynman1982}, quantum computers can directly encode a quantum system and hence efficiently simulate any quantum Hamiltonian dynamics.
With the help of quantum computers, various quantum algorithms have been to proposed~\cite{berry2007efficient,berry2012black,TaylorSeries,PhysRevLett.118.010501,low2019hamiltonian}  and applied for studying different problems such as spin models~\cite{somma2002simulating}, fermionic lattice models~\cite{PhysRevA.92.062318}, quantum chemistry~\cite{PhysRevA.90.022305,babbush2015chemical,babbush2014adiabatic},  and quantum field theories~\cite{jordan2012quantum}.
For a given Hamiltonian $H=\sum_{l=1}^{L} \alpha_l H_l$ with coefficients $\alpha_l$ and Pauli operators $H_l$, we focus on the simulation of the time evolution operator $U_0(t)=e^{-iHt}$, which is crucial for studying both the dynamic and static properties~\cite{sethuniversal,aspuru2005simulated,RevModPhys.92.015003}. 
We consider the product formula methods~\cite{trottersuzuki} and ones based on linear-combination-of-unitaries (LCU)~\cite{childs2012hamiltonian}.
They have shown different asymptotical gate complexities with respect to the simulation time $t$, the accuracy $\varepsilon$, and the property of $H$. 

For the product formula method~\cite{suzuki1991general,Childs9456,childs2019nearly, childs2018faster,campbell2019random,chen2020quantum}, also known as the Trotter-Suzuki algorithm, it divides the evolution time $t$ into small time steps $\delta t$ and approximates each $e^{-iH\delta t}$ via a product of $e^{-i\alpha_lH_l\delta t}$, for example as 
$$e^{-iH\delta t}\approx\prod_l e^{-i\alpha_lH_l\delta t}+\mathcal O(\delta t^2).$$ 
The error term $\mathcal O(\delta t^2)$ is a pessimistic worst-case estimation, and it could be dramatically improved by considering commutative relations between different $H_l$. For example, when all $H_l$ are mutually commutative, the approximation becomes accurate and the error vanishes. We refer to Refs.~\cite{Childs9456,heyl2019quantum,childs2020theory,Tran_2020,chen2020quantum,su2020nearly} for error analyses of product formula methods for more general problems. We note that due to the simplicity of the product formula method, it is widely used in experimental implementations of Hamiltonian simulation with near-term quantum devices~\cite{barends2015digital,PhysRevLett.97.050504,PhysRevA.95.042306}.


It is intuitively straightforward to see that the commutative relation could simplify or ease the complexity of Hamiltonian simulation whereas the anti-commutative relation could make the problem hard. While this seems to be true for the product formula method where Hamiltonians consisting of mutually (anti-)commutative terms have zero (maximal) approximation errors, the intuition breaks down in the LCU based methods. Specifically, we consider the truncated Taylor series algorithm~\cite{TaylorSeries}, which, when expanding the operator $e^{-iH\delta t}$ to the $K$th order, gives
$$e^{-iH\delta t} \approx \sum_{j=0}^K \frac{(\sum_l -i\alpha_l H_l\delta t)^j}{j!} + \mathcal O(\delta t^{K+1}).$$
The first summation term is realized via LCU and the approximation error comes from the truncated Taylor expansion error. It is not hard to see that the commutative relation could not universally reduce the approximation error as it does in the product formula method. Instead, the anti-commutative relation may cause cancellation of the higher order approximation error and hence be exploited to increase the simulation accuracy.

In this work, we investigate the role of anti-commutative relation in Hamiltonian simulation. We show that anti-commutative relation could also simplify the complexity of Hamiltonian simulation. We first use the anti-commutative relation to reduce the the truncated error in the Taylor series method, which lead to a tighter error analysis in contrast to the worse case one. 
Then we propose two modified LCU algorithms that exploit the anti-commutative relation and further reduce high-order truncation errors. Moreover, we show that Hamiltonians with mutually anti-commutative terms are perfectly simulable. 
We numerically test the performances of the proposed methods for electronic Hamiltonians and find  significant improvements of the simulation accuracy with shallower quantum circuits. Our work thus broadens our understanding of commutative and anti-commutative relations in Hamiltonian simulation.


\section{Preliminary}
In this section, we review the product formula method and the Taylor series method. We show how the commutative relation plays an important role in the product formula method. In this work, we assume that the Hamiltonian $H$ is expressed as
\begin{equation}
    H=\sum_{l=1}^{L} \alpha_l H_l,
\end{equation}
where $H_l$ is a tensor product of Pauli operators and $\alpha_l$ are real coefficients. 
Each pair of elements $H_i$ and $H_j$ are either commutative or anti-commutative,
\begin{equation}
\begin{aligned}
&[H_i,H_j]=H_iH_j-H_jH_i=0~(\textrm{commutative}),\\
&\{H_i,H_j\}=H_iH_j+H_jH_i=0 ~(\textrm{anti-commutative}).
\end{aligned}
\end{equation}
The target is to simulate the time evolution operator $U_0(t)=e^{-iHt}$ with 
quantum circuits.

\subsection{Product formula method}

The product formula (PF) approach approximates $U_0(t)$ via the product of exponentials of the individual operators $H_l$. 
By splitting the evolution time $t$ into $r$ segments ($r$ also known as the number of Trotter steps) and making $r$ sufficiently large, we can ensure the simulation is arbitrarily precise. In the first order PF method, we approximate $U_0(t)$ via a product form,
\begin{equation}\label{trotter}
   \begin{aligned}
U_0(t)&=e^{-i\sum_l \alpha_l H_lt}\\
&\approx U_{1}(t)\\
&= \left( \prod_{l=1}^{L} e^{-i \alpha_l H_l t/r} \right) ^r.
\end{aligned} 
\end{equation}
Recent error analyses~\cite{Childs9456,childs2020theory} showed that the approximation error is related to the commutative relation of the operators
\begin{equation}
\begin{aligned}\label{Eq:trotremain}
\left\|U_{1}(t)-U_0(t)\right\|\le  \frac{t^2}{2r}\sum_{l_1=1}^L \bigg\|\sum_{l_2\ge l_1}^{L}[\alpha_{l_2} H_{l_2},\alpha_{l_1}H_{l_1}]\bigg\|,\\
\end{aligned}
\end{equation}
where $\|\cdot\|$ corresponds to the spectral norm. The approximation error becomes smaller when more pairs of $H_{l_1}$ and $H_{l_2}$ commute, and the error vanishes when every term commutes to each other.  Note that any two Pauli operators either commute or anti-commute, and thus the approximation error mainly originates from the anti-commutative relation.

\subsection{Taylor series method}
Now we review the truncated  Taylor series (TS) method~\cite{TaylorSeries}, which is a recently introduced more advanced Hamiltonian simulation algorithm that achieves a different asymptotic error scaling. 
The main idea is to apply the (truncated) Taylor series of the evolution operator
and implement the procedure for LCU. We rewrite the evolution operation in the Taylor series,
$$U_0(t)=e^{-iHt}=\sum_{k=0}^{\infty }\frac{(-itH)^k}{k!},$$ 
and the truncated Taylor series to the $K$-th order is
\begin{equation}
\begin{aligned}\label{Eq:truncated}
\widetilde{U}(t)&=\sum_{k=0}^{K}\frac{(-itH)^k}{k!},\\
&=\sum_{k=0}^{K}\sum_{l_1,\cdots,l_k=1}^L \frac{t^k}{k!} \alpha_{l_1}\cdots\alpha_{l_k} (-i)^k H_{l_1}\cdots H_{l_k},\\
&=\sum_{j=0}^J \beta_j V_j,
\end{aligned}
\end{equation}
which is a linear combination of unitaries $V_j=(-i)^k H_{l_1}\cdots H_{l_k}$ with the coefficients $\beta_j=\frac{t^k}{k!} \alpha_{l_1}\cdots\alpha_{l_k}$. 
Without loss of generality, here we set each coefficient $\alpha_l > 0$ and thus each $\beta_j$ is positive.

We also need to divide evolution time $t$ into $r$ segments. In each segment, the algorithmic error of the approximation mainly comes from the finite truncated Taylor series
\begin{equation}
\begin{aligned}
\|\widetilde{U}(t/r)-U_0(t/r)\|\le e^{\alpha t/r}\frac{(\alpha t/r)^{K+1}}{(K+1)!}.
\end{aligned}
\end{equation}
where $\alpha$ denotes the summation of $\alpha_l$, $\alpha=\sum_{l}\alpha_l $.
Given a fixed total allowed error $\varepsilon$, the truncated order parameter $K$ can be chosen as
\begin{equation}
\begin{aligned}
K=O\left(\frac{\log \frac{\alpha t}{\varepsilon}}{\log \log \frac{\alpha t}{\varepsilon}}\right).
\end{aligned}
\end{equation}
We refer to Appendix~\ref{Appendix:Taylor} and Ref.\cite{TaylorSeries,Childs9456} for details of the implementation and error analysis. 
From the above analysis, the commutative relation does not help much in reducing the truncation error. The truncation error exists even if the Hamiltonian consists of mutually commutative terms (as long as $H\neq 0$). In the next section, we show that it is actually the anti-commutative relation that plays a more important role in determining the error of the Taylor series method. 
We further show that the anti-commutative relation can be applied to reduce the algorithmic error and design modified Taylor series methods.

\section{Anti-commutative cancellation}
We first show the general idea of anti-commutative cancellation in refining the error analysis.
We expand the time evolution operator $U_0(t)=e^{-iHt}$  as
\begin{equation}
    U_0(t) = \sum_{m=0}^{\infty} \frac{(-it)^m}{m!} H^m,
\end{equation}
where
\begin{equation}\label{Eq:Hmexpansion}
\begin{aligned}
    H^m&=\left(\sum_l\alpha_{l}H_l\right)^m\\
    &=\sum_{l_1,\dots, l_{m}=1}^L \alpha_{l_1}\alpha_{l_2}\dots\alpha_{l_m}H_{l_1}H_{l_2}\dots H_{l_m}.
\end{aligned}
\end{equation}
Each $H^m$ has $L^m$ terms and its spectral norm is bound by
\begin{equation}
\|H^m\|\le \left(\sum_{l=1}^{L} \alpha_l \right)^m=\alpha^m.
\end{equation}
In the original error analysis, this upper bound  $\alpha^m$ is used to estimate the truncation error, which is overestimated without considering the relations between different $H_l$. 
To refine the estimation of the norm $\|H^m\|$, we consider each sequence in Eq.~\eqref{Eq:Hmexpansion}  
\begin{equation}
    H_{l_1}H_{l_2}\dots H_{l_{m-1}} H_{l_{m}},
\end{equation}
where $l_1,\dots l_{m}\in \{1,2,\dots,L\}$.
If two terms, for instance $H_{l_1}$ and $H_{l_2}$ are anti-commutative with $\{H_{l_1},H_{l_2}\}=0$, the product with $H_{l_1}H_{l_2}$ at some positions can cancel out the terms with $H_{l_2}H_{l_1}$ at the same positions, for example as
\begin{equation}
  \alpha_{l_1}\alpha_{l_2}\dots\alpha_{l_m}( H_{l_1}H_{l_2}\dots H_{l_{m}}+H_{l_2}H_{l_1}\dots H_{l_{m}})=0. 
\end{equation}
After a given procedure of cancellation, part of $L^m$ terms in $H^m$ can be cancelled out via this anti-commutative relation and we rewrite $H^m$ as the sum of remaining sequences, 
\begin{equation}
    H^m=\sum_{r=1}^{R}\alpha_{l_1^r}\alpha_{l_2^r}\dots\alpha_{l_m^r} H_{l_1^r}H_{l_2^r}\dots H_{l_m^r}.
\end{equation}
where $l_1^r,\dots,l_m^r\in\{1,\dots,L\}$ and $R$ is the number of remaining terms.
For different methods of cancellation, we could have different $R$. For example, if we use the cancellation pair by pair as shown in the following, for a pair-wisely anti-commutative Hamiltonian, $R=L^{\lfloor\frac{m+1}{2}\rfloor}$. 
In order to estimate the norm of $H^m$,
we mainly focus on the sum of the coefficients in the remaining $R$ terms. Thus we define a key parameter $\alpha^{(m)}$ as 
\begin{equation}\label{}
 \begin{aligned}
\alpha^{(m)}~&=\min \sum_{r=1}^{R} \alpha_{l_1^r}\alpha_{l_2^r}\dots\alpha_{l_m^r}\\
    \text{s.t.}~& H^m=\sum_{r=1}^{R} \alpha_{l_1^r}\alpha_{l_2^r}\dots\alpha_{l_m^r}H_{l_1^r}H_{l_2^r}\dots H_{l_m^r},
  \end{aligned}
\end{equation}
with an upper bound of $\|H^m\|$ as
$\|H^m\|\le \alpha^{(m)}$.
Here the minimization is over all the possible cancellations.

In practice, it is generally hard to directly estimate $\alpha^{(m)}$ for a general Hamiltonian. 
Nevertheless, we can repetitively apply a low order cancellation to serve as an upper bound for higher order cancellations,
\begin{equation}\label{}
 \begin{aligned}
\alpha^{(km)}\le (\alpha^{(k)})^m.
\end{aligned}
\end{equation}
Here we take a second order cancellation as an example to show the cancellation procedure.
We first focus on the positions $(1,2)$ in a sequence of $H_l$, and cancel out all the anti-commutative pairs on positions $(1,2)$,
\begin{equation}
\begin{aligned}
  H_{l_1}H_{l_2}\dots H_{l_{m}}&+H_{l_2}H_{l_1}\dots  H_{l_{m}}=0,\\ ~\{H_{l_1}&,H_{l_2}\}=0.
\end{aligned}
\end{equation}
Then all the rest sequences satisfy
\begin{equation}
\begin{aligned}
  H_{l_1}H_{l_2}\dots H_{l_m}, ~(l_1,l_2)\in Comm,
\end{aligned}
\end{equation}
where $Comm=\{(i,j)|[H_i,H_j]=0\}$ denotes the set of commutative pairs and 
$|Comm|$ denotes the number of elements in this set. We then repetitively cancel out the anti-commutative pairs on the positions $(3,4),\dots,(2P-1,2P)$ one by one, where $P=\lfloor\frac{m}{2}\rfloor$. 
At last, the remaining terms satisfy that 
\begin{equation}
\begin{aligned}
    Even~m:~&H_{l_1}H_{l_2}\dots H_{l_{2P-1}} H_{l_{2P}},\\ 
    Odd~m:~& H_{l_1}H_{l_2}\dots H_{l_{2P-1}} H_{l_{2P}}H_{l_{2P+1}},\\
    (l_{2j-1}, l_{2j})&\in Comm,~j=1,\dots,P.\\
\end{aligned}
\end{equation}
Consequently, there are $|Comm|^{\lfloor\frac{m}{2}\rfloor}$, $|Comm|^{\lfloor\frac{m}{2}\rfloor}L$ terms left in $H^m$ for even $m$ and odd $m$, and $H^m$ can be expressed as
\begin{equation}
\begin{aligned}
H^m
=\sum_{(l_{2j-1}, l_{2j})\in Comm}\alpha_{l_1}\dots\alpha_{l_m} H_{l_1}\dots H_{l_m},
\end{aligned}
\end{equation}
Based on the above cancellation,
we obtain the upper bounds for $\alpha^{(m)}$
 \begin{equation}
\begin{aligned}
\alpha^{(2)}  &\le \alpha_{comm},\\
\alpha^{(m)}  &\le \alpha_{comm}^{\lfloor \frac{m}{2}\rfloor},~ even~m,\\
\alpha^{(m)}  &\le \alpha\alpha_{comm}^{\lfloor \frac{m}{2}\rfloor}, ~odd~m,\\
\end{aligned}
\end{equation} 
where $\alpha_{comm}=\sum_{(i,j)\in Comm}\alpha_i\alpha_j$, and $\alpha=\sum_{l=1}^{L}\alpha_l$ denote the summation of the coefficients of commutative pairs and the summation of coefficients of all $L$ terms, respectively.
As a result, the spectral norm of $H^m$ is bounded as
\begin{equation}
\begin{aligned}\label{Eq:Hmupp}
\|H^m\|\le \alpha_{comm}^{\lfloor\frac{m}{2}\rfloor},~ even~m,\\
\|H^m\|\le \alpha\alpha_{comm}^{\lfloor\frac{m}{2}\rfloor},~ odd~m.\\
\end{aligned}
\end{equation}
Similarly, we can directly estimate the third order $\alpha^{(3)}$ and the fourth order cancellation $\alpha^{(4)}$ numerically and use them to bound   $\alpha^{(m)}$.

\section{Applications of anti-commutative cancellation}
In this section, we show various applications of anti-commutative cancellation. We first use it to 
reduce the algorithmic error in the truncated Taylor series method in Sec.\ref{Sec:Recalculate}.
Secondly, we propose a modified LCU algorithm to further reduce the high order error in the truncated Taylor expansion. 
Finally, we explore an extreme case where all the terms are pair-wisely anti-commutative and propose a modified LCU method tailored for this Hamiltonian. Note that the first two applications  are suitable for all Hamiltonians, though different Hamiltonians would have different degree of improvement over the original method.

\subsection{\qi{Tightening} errors in truncated Taylor series algorithm}\label{Sec:Recalculate}
In this section, we show how the above anti-commutative cancellation could be applied to have a tighter estimation of the  algorithmic error in the Taylor series method. 
In the original Taylor series method with expansion up to the $K$-th order, the algorithmic error comes from all the high-order terms~\cite{TaylorSeries,Childs9456}, which is 
\begin{equation}
 \begin{aligned}\label{Eq:orierr}
&\left\|\widetilde{U}(t)-\sum_{k=0}^{\infty}\frac{(-itH)^k}{k!}\right\|=\left\|\sum_{k=K+1}^{\infty}\frac{(-itH)^k}{k!}\right\|\\
&\le \sum_{k=K+1}^{\infty}\frac{(t\sum_{l}\qi{\alpha_l})^k}{k!}\\
&\le \frac{(t\sum_{l}\qi{\alpha_l})^{K+1}}{(K+1)!}\left[\sum_{j=0}^{\infty}\frac{(t\sum_{l}\qi{\alpha_l})^j}{j!} \right] \\
&=\frac{(t\alpha)^{K+1}}{(K+1)!}e^{t\alpha}.
\end{aligned}
\end{equation}
Here we take all the high-order terms into account and the inequality is not tight since some high-order terms may cancel out each other according to the above analysis. 

Now, we give a refined error estimation for the Taylor series method using the above canceling method. We divide the high order terms into two parts: odd order terms and even order terms. And the spectral norm of each part can be bound via Eq.~\eqref{Eq:Hmupp}. 
We denote $\frac{\alpha}{\sqrt{\alpha_{comm}}}=q$ and
the refined error can be expressed as
\begin{equation}
 \begin{aligned}\label{Eq:recalerror}
&\left\|\widetilde{U}(t)-\sum_{k=0}^{\infty}\frac{(-itH)^k}{k!}\right\|\\
&\le \frac{(t\alpha)^{K+1}}{(K+1)!}\frac{(q+1) e^{t\alpha/q}+(-1)^K(q-1)e^{-t\alpha/q}  }{2q^{K+1}}.\\
\end{aligned}
\end{equation}
The key parameter $q=\frac{\alpha}{\sqrt{\alpha_{comm}}}$ is related to the amount of anti-commutative relations in $H$. This is because $\alpha^2=\alpha_{comm}+\alpha_{anti}$ where $\alpha_{anti}=\sum_{i,j\in Anti}\alpha_i\alpha_j$ with $Anti = \{(i,j)|\{H_i,H_j\}=0\}$, and $1/q^2 = \alpha_{comm}/\alpha^2$ thus measures the portion of commutative pairs.
Comparing to the errors in Eq.~\eqref{Eq:orierr}, we reduce it by a factor of $O(\frac{1}{q^{K+1}})$.
If all the terms are commutative with $q=1$, it recovers the previous result. However, when  $H$ is very anti-commutative with a small $1/q$ or a large $q$, the refined error becomes much tighter.
For example, consider a Hamiltonian $H$ with the similar magnitudes of coefficients, and half of the pairs of terms are commutative, then $q$ is roughly $\sqrt{2}$ and we can reduce the error to $2^{-(K+1)/2}$.
For a Hamiltonian  $H$ with pair-wisely anti-commutative properties, we have $\alpha_{comm}=\sum_l \alpha_l^2$, $q=\sqrt{L}$, and the error can be  reduced to $L^{-K/2}$. 
 We give more details about the refined error with third and general $p$ order cancellation in Appendix~\ref{Appendix:reduce}.


\subsection{Modified Taylor series method reducing high order truncated errors} \label{Sec:reduce}
Here in this section, we further present a refined Taylor series method that exploits anti-commutative cancellation and takes account of high order terms with negligible additional cost.
We denote all the high order terms (more than $K-1$) as
\begin{equation}
\begin{aligned}
R_{K}&=\frac{(-itH)^{K}}{K!}+\sum_{k=K+1}^{\infty}\frac{(-itH)^{k}}{k!}\\
&=\frac{(-itH)^{K-1}}{K!}\left( -itH+\sum_{j=2}^{\infty}\frac{(-itH)^{j}}{(K+j-1)!/K!}\right)\\
&=\frac{(-itH)^{K-1}}{K!}\left[ \sum_{l=1}^L{\alpha_l}(-i tH_l) + \sum_{\vec{e}} \widetilde{\gamma_{\vec{e}}}  H_{\vec{e}} \right].\\
\end{aligned}
\end{equation}
Here in the last line, we rewrite the terms $-iHt$ and $\sum_{j=2}^{\infty}\frac{(-itH)^{j}}{(K+j-1)!/K!}$ as $\sum_{l=1}^L{\alpha_l}(-i tH_l)$, $ \sum_{\vec{e}} \widetilde{\gamma_{\vec{e}}} H_{\vec{e}} $, respectively. 
Here $H_{\vec{e}}$ is the product of $H_l$ that remains after a certain canceling procedure which can be expressed as
\begin{equation}
\begin{aligned}
H_{\vec{e}}=(-i)^k H_{e_1}H_{e_2}\dots H_{e_k}.
\end{aligned}
\end{equation}
In the original truncated Taylor LCU method, all these $H_{\vec{e}}$ are discarded as error terms.   
However, some $H_{\vec{e}}$ may equal to simple unitaries, such as identity $I$, or unitaries that appears in lower order expansions, such as $H_l$.
For example, according to the anti-commutative cancellation, the second and third order terms $H^2$ and $H^3$ can be expressed as
\begin{equation}
 \begin{aligned}\label{Eq:H2H3}
H^2&=  \beta_0 I+\sum_{(l_1,l_2)\in Comm^*} \alpha_{l_1}\alpha_{l_2}H_{l_1}H_{l_2},\\
H^3&=\sum_l \beta_l H_l +\\
&\sum_{(l_1,l_2),(l_1,l_3),(l_2,l_3)\in Comm^*} \alpha_{l_1}\alpha_{l_2}\alpha_{l_3}H_{l_1}H_{l_2}H_{l_3},
\end{aligned}
\end{equation}
where we define the set containing all the commutative pairs without the same terms as 
$Comm^*=\{(i,j)|[H_i,H_j]=0,~ i\neq j\}$. 
Here we use the facts $H_l^2=I$ and $H_lH_{l'}H_{l'}=H_l$. 
For high order terms, the following sequences equal to $I$ and $H_l$,
\begin{equation}
\begin{aligned}
   & Even~length : (H_{l_1}H_{l_1})\dots (H_{l_{P}} H_{l_{P}})=I, \\
    &Odd~length:  (H_{l_1}H_{l_1})\dots (H_{l_{P}} H_{l_{P}})H_{l} =H_l.\\
\end{aligned}
\end{equation}
Instead of discarding all these terms, we can update the coefficients of $-iH_l$ from $\alpha_lt$ to a new one $\gamma_l$ that includes higher order contributions and add the term $U_0=-I$. 
Moreover, we can add other unitaries with the modified coefficients apart from the unitaries $-iH_l$ and $U_0=-I$.
For example, we choose $E$ number of unitaries with the largest coefficients in the rest of $H_{\vec{e}}$, denoted as $U_1$ to $U_E$. 
Consequently, we construct an approximate $K$-th order term as
\begin{equation}
\begin{aligned}\label{Eq:Rk}
\widetilde{R}_{K}
&=\frac{(-itH)^{K-1}}{K!}\left[ \sum_{l=1}^L{\gamma_l}(-i H_l) + \sum_{j=0}^{E} \widetilde{\gamma_{j}} U_{j} \right],\\
\end{aligned}
\end{equation}
and we can implement 
\begin{equation}
 \widetilde{U}(t)=  \sum_{k=0}^{K-1}\frac{(-itH)^k}{k!} +\widetilde{R}_{K}
\end{equation}
via the LCU routine. Specifically, we implement the low order (less than $K-1$) terms as usual and the $K$-th term $\widetilde{R}_{K}$ with a modified circuit that involves little additional resource. The details of additional gates are shown in the numerical section. Overall, the refined method could reduce the error with only a few additional gates. It works well for Hamiltonians with large anti-commutative terms or very biased coefficients. 


Now we analyze the error for the modified scheme. 
We divide the error into three parts $(K+1)$-th order error $E_{K+1}$, $(K+2)$-th order error $E_{K+2}$, and high order error $E_{\infty}$ as
\begin{equation}
 \begin{aligned}
&\left\|\widetilde{U}(t)-\sum_{k=0}^{\infty}\frac{(-itH)^k}{k!}\right\|=\|E_{K+1}+E_{K+2}+E_{\infty}\|.\\
\end{aligned}
\end{equation}
Without loss of generality, we assume $K$ is odd. Benefiting from anti-commutative relation, we could obtain a more accurate analysis of $E_{K+1}$
and $E_{K+2}$ which take up a large proportion of the total error and are bound as
\begin{equation}
 \begin{aligned}\label{Eq:modierror2}
&\|E_{K+1}\|\le\frac{t^{K+1}\alpha_{comm}^{ \frac{K-1}{2}} E_{\epsilon}}{(K+1)!},\\
&\|E_{K+2}\|\le \frac{t^{K+2}\alpha_{comm}^{ \frac{K-1}{2}}\alpha^{(3)}_r }{(K+2)!},\\
\end{aligned}
\end{equation}
where $E_{\epsilon}$ is the sum of 
coefficients of the remaining terms 
in $\sum_{(l_1,l_2)\in Comm^* } \alpha_{l_1}\alpha_{l_2}H_{l_1}H_{l_2}$ in Eq.~\eqref{Eq:H2H3}, which are not included in $U_{j}$ and $-iH_l$ and $E_{\epsilon}\le \alpha_{comm}-(\sum_l \alpha_l^2)$, and 
$\alpha^{(3)}_r$ is the sum of coefficients $\alpha_{l_1}\alpha_{l_2}\alpha_{l_3}$ in $H^3$ according to Eq.~\eqref{Eq:H2H3}.
We could directly use the bound in 
Eq.~\eqref{Eq:recalerror} for high order terms $\|E_{\infty}\|$.
Though we could also calculate the sequences which equals to $H_l$ or $I$ in $E_{\infty}$, it would not influence the results a lot.


With the parameter $\frac{\alpha}{\sqrt{\alpha_{comm}}}=q$, 
 the error can be expressed as
\begin{equation}
 \begin{aligned}\label{Eq:modierror}
&\left\|\widetilde{U}(t)-\sum_{k=0}^{\infty}\frac{(-itH)^k}{k!}\right\|\\
&\le \frac{(t\alpha)^{K+1} }{(K+1)!q^{K-1}}\frac{E_{\epsilon}}{\alpha^2}+ \frac{(t\alpha)^{K+2}}{(K+2)!q^{K-1}}\frac{\alpha^{(3)}_r}{\alpha^3}\\
&+\frac{(t\alpha)^{K+3}}{(K+3)!}\frac{(q+1) e^{t\alpha/q}+(-1)^K(q-1)e^{-t\alpha/q}  }{2q^{K+3}}.
\end{aligned}
\end{equation}
We refer to Appendix 
\ref{Appendix:Taylor} and \ref{App:modif}
for details of the implementation
and error analysis.
From the above expression, we know that this error is related to some key parameters $\alpha_{comm}$, $E_{\epsilon}$ and $\alpha^{(3)}_r$. 
Though this modified method could not influence  asymptotic performance, it could significantly reduce the error in the near term  implementation of the Taylor series method.
The numerical verification of the improvement on performance is shown in the  next section.

\subsection{Modified Taylor series method with pair-wisely anti-commutative terms}\label{Sec:perfect}
At last, we consider an extreme case with Hamiltonians consisting of pair-wisely anti-commutative terms. 
For example, we construct an $n$-qubit Hamiltonian $H=\sum_{j=1}^{n} H_j$ with $n$ pair-wisely anti-commutative terms,
\begin{equation}
    \begin{aligned}
&H_1=X_1, H_2=Z_1\otimes Z_2\\ 
&H_j=Z_1\otimes X_2\dots \otimes X_{j-1} \otimes Z_j, j=3,\dots,n.
    \end{aligned}
\end{equation}
In this section, we show how the evolution of this kind of Hamiltonians can be simulated via the LCU method without any approximation error. 
We first show the exact form of the expansion of the Hamiltonian.
\begin{lemma}\label{Le:perfect}
For a Hamiltonian $H=\sum_{l=1}^L \alpha_l H_l$ where all the terms $H_l$ are unitary and mutually anti-commutative,
$\{H_{l_1},H_{l_2}\}=0$, $l_1\neq l_2$,
$H^m$ can be expressed as
\begin{equation}
\begin{aligned}
  H^m&=\gamma^{(m)}_0 I +\sum_{l=1}^{L}\gamma^{(m)}_l H_{l},
\end{aligned}
\end{equation}
for some coefficients $\gamma^{(m)}_l$, $l=0,\dots,L$.
\end{lemma}
\noindent In particular, we have 
\begin{equation}
    \begin{aligned}
&\gamma^{(2j-1)}_0 = 0, \,\gamma^{(2j-1)}_l =\beta_s^{2j-2} \alpha_l,\\
&\gamma^{(2j)}_0 = \beta_s^{2j}, \,\gamma^{(2j)}_l = 0, 
    \end{aligned}
\end{equation}
where  $j\in \mathbb N^+$ is a positive integer and 
$\beta_s=\sqrt{\sum_{l=1}^{L}\alpha_l^2}$.
Now, we expand the evolution $U(t)=e^{-iHt}$, and we can rewrite it as
\begin{equation}
\begin{aligned}\label{Eq:perfectU}
U(t)=\widetilde{U}(t)&= \tilde{\alpha}_0 I+ \sum_{l} \tilde{\alpha}_l (-iH_l),\\
\end{aligned}
\end{equation}
where  
\begin{equation}
\begin{aligned}
\tilde{\alpha}_0&=1-\beta_s^2t^2/2!+\beta_s^4t^4 /4!+\dots=\cos (t\beta_s),\\
\tilde{\alpha}_l&=\alpha_lt-\alpha_l\beta_s^2t^3/3!+ \alpha_l\beta_s^4 t^5/5!+\dots\\
&=\frac{\alpha_l}{\beta_s} \sin(t\beta_s), 
~l=1,\dots,L.
\end{aligned}
\end{equation}
Note that here the coefficients $\tilde{\alpha}_l$ may not be always positive, and we could replace them with the absolute values and change the sign of the unitary gate accordingly. 
The sum of the absolute value coefficients is 
\begin{equation}
\begin{aligned}\label{Eq:perefects}
    s&=\sum_{l}|\tilde{\alpha}_l|=|\cos (t\beta_s)|+\frac{\alpha}{\beta_s}|\sin (t\beta_s)|,
\end{aligned}    
\end{equation}
which is a function of $t$ and $\beta_s$.  
In particular, we have $L\beta_s^2\ge \alpha^2$ and thus $1\le s\le 1+\sqrt{L}$. 
\begin{theorem}
For a Hamiltonian $H=\sum_{l=1}^L \alpha_l H_l$ acting on $n$ qubits where all the terms $H_l$ are mutually anti-commutative,
$\{H_{l_1},H_{l_2}\}=0$, $l_1\neq l_2$, we can implement the evolution $U(t)=e^{-iHt}$ without algorithmic error via the LCU method with the gate complexity $\mathcal O(L^{\frac{3}{2}}(n+\log L))$.
\end{theorem}
Here the gate complexity $\mathcal O(L^{\frac{3}{2}}(n+\log L))$ is independent of $t$ and $\varepsilon$. The details of the implementation of LCU is shown in Appendix~\ref{Sec:LCUimple}.
Moreover, the expansion of the unitary also indicates efficient classical simulation of the Hamiltonian similar to Hamiltonians with mutually commutative terms. 
We can also extend the above strict case to more general cases.
\begin{corollary}
Suppose there exists a positive integer $M$, satisfying that $H^M=\gamma I$, we can implement the evolution $U(t)=e^{-iHt}$ without algorithmic error via the LCU method.
\end{corollary}
\noindent Similarly, 
we rewrite the unitary as
\begin{equation}
\begin{aligned}
U(t)&=\sum_{k=0}^{\infty }\frac{(-itH)^k}{k!}
= \sum_{k=0}^{M-1 }\sum_{j=0}^{\infty } (-itH)^k\frac{(-it)^{jM}\gamma^j}{(k+jM)!},\\
&= \sum_{k=0}^{M-1 } \gamma_k(-itH)^k,
\end{aligned}
\end{equation}
where coefficients $\gamma_k=\sum_{j=0}^{\infty } \frac{(-it)^{jM}\gamma^j}{(k+jM)!}$. We can implement this unitary like the original truncated Taylor method with at most $M-1$ order without any algorithmic error.
In the above analysis, 
we require that the Hamiltonians are mutually anti-commutative, which is rather strict.
\begin{corollary}\label{Cor:nearp}
When the Hamiltonian $H$ is not perfectly anti-commutative, but close to the perfect case,\begin{equation}
\begin{aligned}\label{Eq:epsiA}
    H^2=\beta_s^2 I +V, ~~\|V \|\le \varepsilon_{A}.
\end{aligned}
\end{equation}
There exist coefficients $\tilde{\alpha}_l$, such that
$\widetilde{U}(t)$ is close to $U_0(t)=\sum_{k=0}^{\infty}\frac{(-itH)^k}{k!}$ with an error $\mathcal O(t^2\varepsilon_A)+\mathcal O(t^3\varepsilon_A\alpha)$ for a small $t$,
\begin{equation}
\begin{aligned}
&\widetilde{U}(t)= \tilde{\alpha}_0 I+ \sum_{l} \tilde{\alpha}_l (-iH_l),\\
&\left\|\widetilde{U}(t)-\sum_{k=0}^{\infty}\frac{(-itH)^k}{k!}\right\|=\mathcal{O}(t^2\varepsilon_A)+\mathcal O(t^3\varepsilon_A\alpha).\\
\end{aligned}
\end{equation}
\end{corollary}
Note that here we need a small time $t$. Thus for a long time $t$, we could also divide it into $r$ segments, and the total error is $\mathcal{O}(t^2\varepsilon_A/r)+\mathcal O(t^3\varepsilon_A\alpha/r^2)$. When $t=n$, and $\alpha=\mathcal O(n)$, then the error is $\mathcal{O}(n^2\varepsilon_A/r)$.
The parameter for commutative pairs has the upper bound, 
$\varepsilon_A\le \sum_{(i,j)\in Comm^*}\alpha_i\alpha_j< \alpha_{comm}$.
Compared to the error in the first order PF $\mathcal{O}(n^2\alpha_{anti}/r)$, the differences are the prefactors quantifying the 
anti-commutation and commutation property. When the Hamiltonian is more anti-commutative, $\alpha_{comm}< \alpha_{anti}$,
then this approximate LCU method could has the advantage over the first order PF.

\section{Numerical simulation}
Here in this section, we show the numerical performance of the proposed methods. 
Lattice Hamiltonians mostly consist of low weight terms, which are likely to be commutative to each other; whereas chemistry Hamiltonians, after the fermion-to-qubit encoding, have more anti-commutative terms.
Since the proposed methods are more suitable for an anti-commutative Hamiltonian, we focus on the electronic structure of chemistry Hamiltonians. 
The second quantized form of the electronic Hamiltonians is
\begin{equation}
H=  \sum_{p,q}h_{pq} a_p^{\dagger}a_q+\frac{1}{2}\sum_{p,q,r,s}h_{pqrs} a_p^{\dagger}a_q^{\dagger}a_r a_s,
\end{equation}
where $a^{\dagger}$ and $a$ are creation and annihilation  operators \cite{RevModPhys.92.015003}. 
Then it is converted to a qubit Hamiltonian via the Jordan-Wigner transformation \cite{jordan_uber_1928}. 
\begin{equation}
\begin{aligned}
a_p&=\frac{X_p+iY_p}{2}\otimes  Z_{p-1} \otimes \dots\otimes Z_0,    \\
a_p^{\dagger}&=\frac{X_p-iY_p}{2}\otimes  Z_{p-1} \otimes \dots\otimes Z_0.
\end{aligned}
\end{equation}
We consider 23 different molecules as shown in Table~\ref{Table:second} and the Hamiltonian data is from Ref.~\cite{McClean_2020}. 

\subsection{Refined errors estimation}
In the Taylor series method, the total time is divided into several segments with each time segment $t=\ln2/\alpha$.
The error in the original analysis is
\begin{equation}
\begin{aligned}\label{Eq:ori}
\varepsilon_{o}=\frac{\delta_{o}^2+3\delta_{o}+4}{2}\delta_{o},~
\delta_{o}=2\frac{(\ln2)^{K+1}}{(K+1)!}.
\end{aligned}
\end{equation}
where $\delta_o$ is the truncated error in the Taylor expansion. 
According to Eq.~\eqref{Eq:recalerror},
we calculate the refined errors $\varepsilon_{n}$ and show the improvement of the refined error estimation for various molecules 
with up to 25 qubits.
In order to show the improvement clearly, we calculate the ratios ${\varepsilon_{o}}/{\varepsilon_{n}}$
versus different truncated order $K=10, 20, 30, 40$ in Table \ref{Table:second} using the second order cancellation.
\begin{table}[htb]
\scalebox{0.75}{
\begin{tabular}{lllllll}
\hline
\hline
\multicolumn{7}{c}{Error Reduction $\varepsilon_o/\varepsilon_n$  }\\
\hline
Formula &Qubits &L& $\rm K=10$        & $\rm K=20$  &$\rm K=30$ &$\rm K=40$   \\
\hline
HO & 11 & 631   &    1.445&	2.016&	2.813	&3.926   \\
LiH & 11& 631   &  1.866	&3.285	&5.782&	10.177\\
BH & 11 &631  & 1.962	&3.615&	6.660	&12.270 \\
$\rm BeH_2$& 13 &666     & 1.990	&3.714	&6.930	&12.933
  \\
$\rm NH_2$&13 &1086 &  1.655	&2.611	&4.119&	6.500 \\
$\rm BH_2$ & 13 &1086          & 2.111&	4.157&	8.187&	16.123
 \\
$\rm CH_3$ &15 &1969        & 1.919&	3.466	&6.260&	11.305
 \\
$\rm NH_3$ &15 &2929              & 1.806&	3.085	&5.272&	9.007
  \\
$\rm CH_4$ & 17 &6892              & 2.492&	5.708&	13.074	&29.948
  \\
NO & 19 &4427           &  1.428	&1.972&	2.724&	3.761
 \\
CN & 19& 5851         &  1.561	&2.337	&3.498&	5.237
 \\
BN &19 &5851            & 1.652	&2.604&4.105&	6.469\\
LiOH &21& 8750            &  1.639	&2.565	&4.013	&6.278
  \\
HBO & 21& 8758          &  1.663&	2.637&	4.180&	6.627\\
HOF & 21& 12070        & 1.525	&2.234	&3.273	&4.796 \\
CHF &21 &12074        &  1.616&	2.497&	3.857&	5.958 \\
$\rm H_2NO$ &23 &9257       & 1.532	&2.254&	3.317	&4.881
 \\
$\rm CH_2O$ &23 &9257             & 1.582&	2.396&	3.630	&5.499
  \\
$\rm NH_2F$  & 23 &15673       & 1.574	&2.374&	3.581&	5.401
 \\
$\rm CH_2F$ & 23 &15681       &  1.605&	2.465	&3.785	&5.811
 \\
$\rm CH_3F$ &25& 18600        &  1.575	&2.375&	3.583&	5.405
  \\
$\rm CH_3Li$ &25& 19548    &2.014	&3.799	&7.166	&13.518
 \\
$\rm OCH_3$ &25 &39392       & 1.705&	2.763	&4.480 &7.263
 \\
 \hline
\end{tabular}}
\caption{Error Reduction using second-order cancellation $\varepsilon_o/\varepsilon_n$ for various molecules.}\label{Table:second}
\end{table}
\begin{figure}[t]
\centering
\includegraphics[width=.5\textwidth]{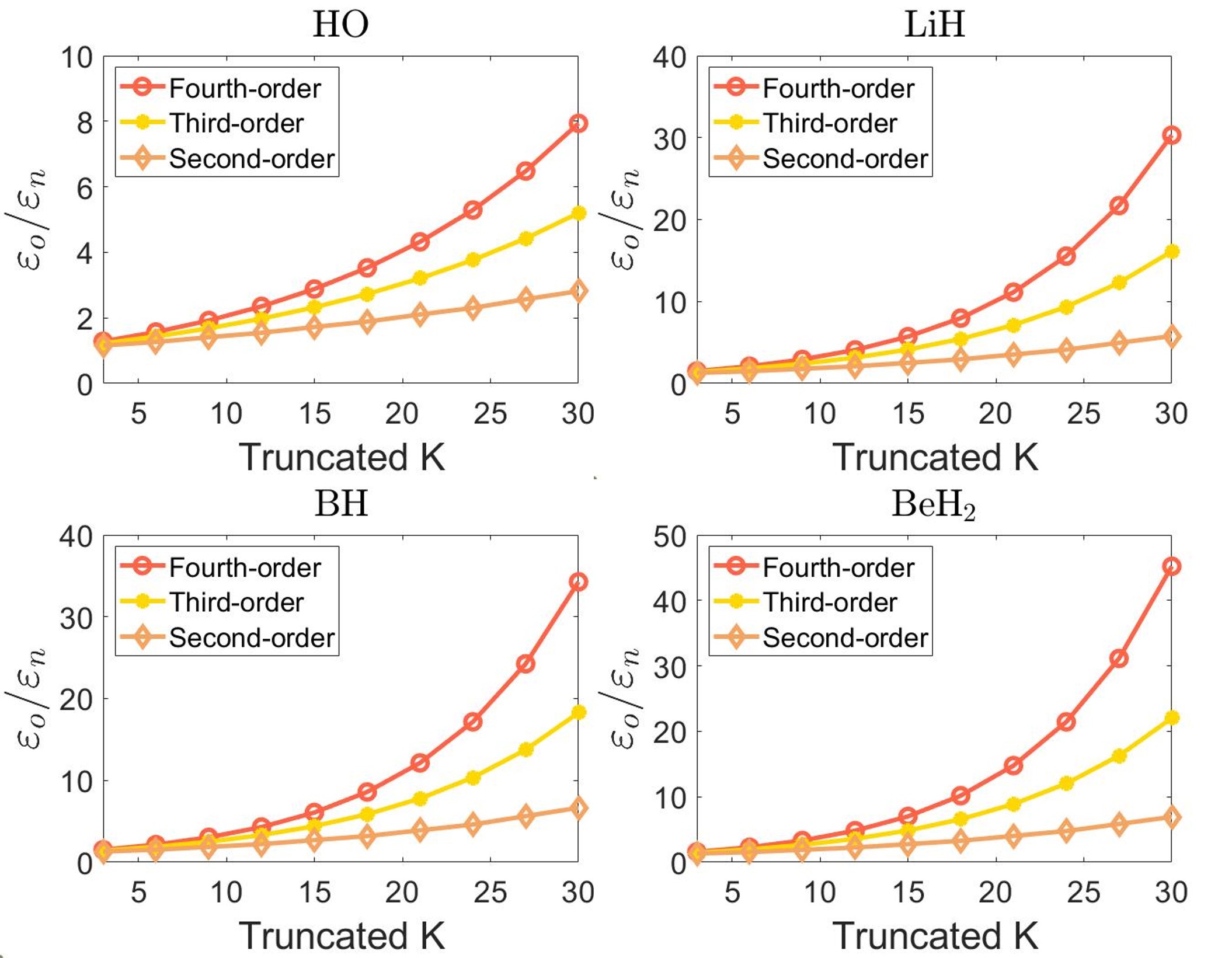}
\caption{ Ratio $\varepsilon_{o}/\varepsilon_{n}$ with second order, third order, and fourth order cancellation versus different truncated order $K$ with $t=\ln2/\alpha$ for molecules $\rm HO, \rm LiH, \rm BH, \rm BeH_2$.} \label{Fig:erro1}
\end{figure}
Moreover, we also plot the ratio ${\varepsilon_{o}}/{\varepsilon_{n}}$
for some example molecules, $\rm HO$, $\rm LiH$, $\rm BH$, and $\rm BeH_2$ using second order cancellation, third order, and fourth order cancellation in Fig.~\ref{Fig:erro1} using the upper bounds in Eq.~\eqref{Eq:p-Newerror}.
The lower bounds of the key parameters $q=\frac{\alpha}{(\alpha^{(2)})^{1/2}}, \frac{\alpha}{(\alpha^{(3)})^{1/3}},\frac{\alpha}{(\alpha^{(4)})^{1/4}}$ can be estimated numerically, which are summarized in Table \ref{Table:234}. 
Though higher order cancellation brings more significant improvement, 
the numerical estimation of high order cancellation also becomes much
harder with the increase of the number of terms $L$ and the number of qubits $n$.  
\begin{table}[htb]
\begin{center}
\scalebox{0.9}{
\begin{tabular}{l|lll}
\hline
\hline
Molecule &  $\frac{\alpha}{(\alpha^{(2)})^{1/2}}$& $\frac{\alpha}{(\alpha^{(3)})^{1/3}}$&$\frac{\alpha}{(\alpha^{(4)})^{1/4}}$\\
\hline
$\rm HO$ &1.0339	&1.0552	&1.0699\\
$\rm LiH$& 1.0582&1.0949&1.1177\\

$\rm BH$ &1.0630&1.0994&1.1222\\
$\rm BeH_2$& 1.0644&	1.1061&1.1324\\

\hline
\end{tabular}}
\end{center}
\caption{
Parameters $\frac{\alpha}{(\alpha^{(2)})^{1/2}}, \frac{\alpha}{(\alpha^{(3)})^{1/3}},\frac{\alpha}{(\alpha^{(4)})^{1/4}}$ 
estimation for second order, third order, and fourth order cancellation.}\label{Table:234}
\end{table}

\subsection{Reducing the high order error in modified algorithm}
In Section~\ref{Sec:reduce}, we further reduce the high order truncated error 
via modifying the coefficients and involving more unitary operations in the Taylor series. 
We calculate three different types of new errors with $t=\ln2/\alpha$: 
\begin{enumerate}
    \item Refined errors shown in Eq.~\eqref{Eq:recalerror} using second order cancellation; 
    \item Reduced errors in modified method without additional unitaries $(E=0)$ in Eq.~\eqref{Eq:modierror}. 
    \item Reduced errors in modified method with $E=2^w-L-1 $ additional unitaries in Eq.~\eqref{Eq:modierror} where $w= \lceil \log_2 L\rceil$.
\end{enumerate}
According to Appendix~\ref{App:modif} and Ref.~\cite{Childs9456}, these three schemes we applied in the numerical simulation have 
almost the same gate cost.
In order to show the improvement clearly, we show the curves of the ratio of the original error to new calculated error, $\varepsilon_{o}/\varepsilon_{n}$, versus different truncated order $K$ for molecules, $\rm BH, \rm CH_4, \rm HBO, \rm CH_3Li$ 
with terms $L$ ranging from $631$ to $19548$. 
We find that with increasing $K$, 
the reduced error becomes more significant. Note that here the improvement is not related to the system size $n$ nor the number of terms $L$, but related to the anti-commutative property of different molecules. 
\begin{figure}[t]
\centering
\includegraphics[width=0.5\textwidth]{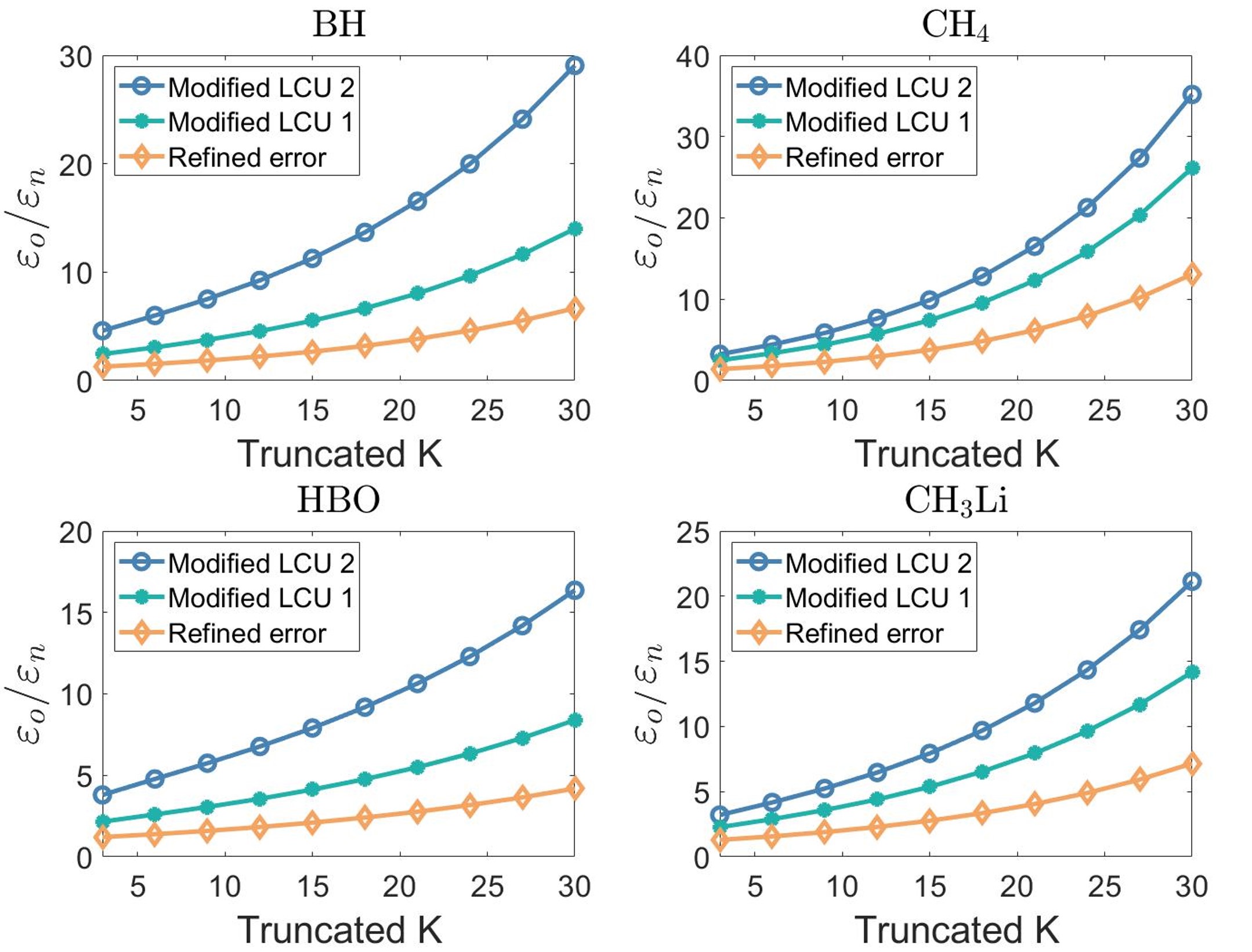}
\caption{ Ratio $\varepsilon_{o}/\varepsilon_{n}$ versus different truncated order $K$ with $t=\ln2/\alpha$ for molecules  $\rm BH, \rm CH_4, \rm HBO, \rm CH_3Li$. Modified LCU 1 and 2 denote the results with $E=0$ and 
$E=2^w-L-1 $ additional unitaries in Eq.~\eqref{Eq:modierror}, respectively.
}\label{Fig:erro2}
\end{figure}


In some Hamiltonian simulation tasks, a predetermined simulation accuracy $\varepsilon$ is given. The task is to find the most efficient circuit approximating the ideal evolution with at most $\varepsilon$ error.
Thus, we also compare the required truncated $K$ with a given simulation accuracy $\varepsilon$ for molecules $\rm BH, \rm CH_4, \rm HBO, \rm CH_3Li$. We choose the simulation time $t=n$ where $n$ is the number of qubits. 
We find that in most of cases shown in Table~\ref{Table:K}, the refined error scheme or the modified scheme could reduce one order of Taylor expansion. 
\qi{Although this does not seem a significant change in theory, it can indeed reduce both the  gate count and the number of ancillary qubits, making its practical implementation much easier. For a Hamiltonian with $L$ unitaries, one order reduction could reduce $\log L$ number of ancillary qubits. We show the total CNOT gates and ancillary qubits reduction for these examples in Table 4. For example, consider a large molecule Hamiltonian with $N=100$ spin orbitals and $h = 50$ electrons, we have about $L\approx N^2 h^2 = 2.5\times 10^7$ terms and in one segment our method can roughly reduce $25$ qubits and accordingly remove $7.5\times 10^8$ CNOT gates that act on those qubits. 
}

\begin{table}[t]
\scalebox{0.7}{
\begin{tabular}{l|lll|lll|lll|lll}
\hline
\hline
\multirow{2}*{Error $\varepsilon$}& \multicolumn{12}{c}{Minimum $K$}
\\
\cline{2-13}
& \multicolumn{3}{c}{$\rm BH$}  & \multicolumn{3}{c}{$\rm CH_4$} & \multicolumn{3}{c}{$\rm HBO$}& \multicolumn{3}{c}{$\rm CH_3Li$} \\
\hline
1e-06 & 10     & 10    & 10       & 11     & 11     & 10  & 11     & 11     & 11   & 11      & 11     & 10     \\
1e-07 & 11     & 11    & 10      & 12     & 11     & 11  & 12     & 12     & 11    & 12      & 12     & 11     \\
1e-08 & 12     & 12    & 11      & 12     & 12     & 12  & 13     & 13     & 12    & 13      & 12     & 12     \\
1e-09       & 13     & 13    & 12       & 13     & 13     & 13   & 14     & 13     & 13  & 13      & 13     & 13     \\
 1e-10          & 14     & 13    & 13      & 14     & 14     & 13   & 14     & 14     & 14   & 14      & 14     & 14     \\
 1e-11           & 14     & 14    & 14        & 15     & 14     & 14 & 15     & 15     & 14   & 15      & 15     & 14     \\
  1e-12          & 15     & 15    & 14       & 15     & 15     & 15  & 16     & 16     & 15   & 16      & 15     & 15     \\
 1e-13           & 16     & 15    & 15    & 16     & 16     & 15   & 16     & 16     & 16     & 16      & 16     & 16     \\
 1e-14           & 16     & 16    & 16      & 17     & 16     & 16   & 17     & 17     & 17   & 17      & 17     & 16     \\
       1e-15     & 17     & 17    & 16       & 18     & 17     & 17   & 18     & 18     & 17  & 18      & 17     & 17     \\
         1e-16   & 18     & 18    & 17      & 18     & 18     & 18  & 19     & 18     & 18    & 18      & 18     & 18     \\
         1e-17   & 19     & 18    & 18      & 19     & 19     & 18  & 19     & 19     & 19    & 19      & 19     & 18     \\
  1e-18        & 19     & 19    & 18     & 20     & 19     & 19  & 20     & 20     & 19     & 20      & 19     & 19     \\
 1e-19          & 20     & 20    & 19       & 20     & 20     & 20  & 21     & 20     & 20   & 21      & 20     & 20     \\
 1e-20          & 21     & 20    & 20       & 21     & 21     & 20 & 21     & 21     & 21    & 21      & 21     & 20    \\
 \hline
\end{tabular}}
\caption{Minimum $K$ for required accuracy in three different schemes for molecules $\rm BH$, $\rm CH_4$, $\rm HBO$, $\rm CH_3Li$. The first column for each molecule is with original Taylor series method.
The second column is with refined error scheme in Sec.~\ref{Sec:Recalculate}. And the third column is with modified scheme in Sec.~\ref{Sec:reduce} with $E=2^w-L-1 $.}
\label{Table:K}
\end{table}

\begin{table}[t]
\qi{
\begin{center}
\scalebox{0.75}{
\begin{tabular}{|l|c|c|c|c|}
\hline
\hline
Reduction& $\rm BH$  &$\rm CH_4$ & $\rm HBO $  &$\rm CH_3Li$. \\
\hline
CNOT Gate &  $8.7\times 10^6 $ &$3.0\times 10^8 $ & $1.5\times 10^9$ &  $2.3\times 10^9$  \\
\hline
Ancillary qubits & 10    &13  &  14   &  15\\
 \hline
\end{tabular}}
\end{center}}
\caption{\qi{Number of CNOT gates and ancillary qubits reduction for one order reduction of Taylor expansion for simulating molecules $\rm BH$, $\rm CH_4$, $\rm HBO$, $\rm CH_3Li$ with simulation time $t=n$ where $n$ is the number of qubits.}}
\label{Table:gatereducation}
\end{table}

\section{Discussion}
In this work, we explore the anti-commutative relation in Hamiltonian simulation. We find that Anti-commutative relation plays a positive role in the truncated Taylor series method.
We show that the simulation accuracy could be significantly improved by utilizing the anti-commutative relation.  
In particular, pair-wisely anti-commutative Hamiltonians admit an efficient and error free simulation with the LCU method.
\qi{Our work is the first attempt of exploiting anti-commutative relations in Hamiltonian reduction, which may inspire and motivate more interesting and significant future works to find more efficient reduction mechanisms that combine with other techniques or considering specific Hamiltonians that have large anti-commutative terms. }

\qi{When we consider some specific types of Hamiltonians, the performance of Taylor series method could be further improved.  For instance, combing our result with techniques in Ref.~\cite{meister2020tailoring},
we may further reduce the simulation error for electronic structure Hamiltonians. 
Here most of our results are based on a low order anti-commutative cancellation. In general, high order anti-commutative cancellation is classically intractable. However, we may find explicit and more efficient strategies to exploit the high order anti-commutative cancellation for some specific Hamiltonians.}

Another direction is exploring the role of anti-commutation in other Hamiltonian simulation algorithms, e.g. product formula methods and quantum signal processing algorithm. From our current understanding, it might seem unlikely that anti-commutative relation could help in a standard product formula methods due to the lack of anti-commutation cancellation. But it might be helpful for the variant of product formula methods, e.g. randomized product formula, which involves the summation of the products of different sequences \cite{childs2018faster,campbell2019random,chen2020quantum}. Meanwhile, due to the complementary advantages of commutation in product formula and anti-commutation in the Taylor series method, we could also consider hybrid approaches of Trotter and Taylor series methods. In this case, the Hamiltonian could be divided into commutative and anti-commutative terms (sets), which are simulated using Trotter and Taylor expansion respectively. Of course, how to divide the terms and whether such a hybrid approach is useful seem a very interesting future direction.

\qi{The error reduction and speed-up in this work mainly come from representing the power of Hamiltonian $H$ with the summation of fewer unitary terms. Thus, the cancellation idea is not restricted to anti-commutative cancellation. For instance, if the Hamiltonian $H$ has certain symmetry properties, then any order of $H$ could be expressed in the summation of the terms in the symmetry subspace, which might be useful for designing more efficient Hamiltonian simulation algorithms.}

\qi{Note that the perfect anti-commutation case admits an efficient classical simulation of Hamiltonian evolution. And the Trotter formula, for instance the second order trotter method is used in the classical algorithms 
of estimating the partition function~\cite{bravyi2015monte,bravyi2017polynomial}. We expect the anti-commutation property may inspire better classical simulation method of quantum problems.   }


\section{Acknowledgement}
We are grateful to Andrew Childs, Di Fang, Lin Lin, Yuan Su, and You Zhou for useful discussions and anonymous referees for their comments on an earlier draft. 
We thank Richard Meister for providing the molecular Hamiltonian data.
QZ acknowledges the support by the Department of Defense through the Hartree Postdoctoral Fellowship at QuICS. 
XY acknowledges support from the Simons Foundation. 

\onecolumngrid
\appendix

\begin{appendix}

\section{Taylor expansion method}\label{Appendix:Taylor}

In order to implement $\widetilde{U}(t)$, we need three key elements: a state preparation procedure accomplished by oracle $G$, a reflection denoted as $R$, and an oracle selecting unitary operations denoted as ${\rm select}(V)$.
We assume that the ancillary system is a $d$ dimensional Hilbert space $\mathcal{H}_{an}$ and the evolution operation $e^{-i Ht}$ acts on an $n$ qubit Hilbert space $\mathcal{H}_s$ with dimension $2^n$.  

Here the oracle $G$ is a unitary that transforms the ancillary state $\ket{0}$ to
\begin{equation}
 \begin{aligned}
G\ket{0}=\frac{1}{\sqrt{s}}\sum_{j}\sqrt{\beta_j}\ket{j},
\end{aligned}
\end{equation}
where $s$ is the normalized parameter,
\begin{equation}
 \begin{aligned}
s=\sum_{j=0}^J \beta_j= \sum_{k=0}^K\frac{(t\sum_{l}a_l)^k}{k!}.
\end{aligned}
\end{equation}
The ${\rm select}(V)$ is defined as
\begin{equation}
\begin{aligned}
{\rm select}(V)=\sum_{j} \ket{j}\bra{j} \otimes V_j.
\end{aligned}
\end{equation}
The control register $\ket{j}\bra{j}$ acts on the ancillary system $\mathcal{H}_{an}$.
We construct an operator as
\begin{equation}
 \begin{aligned}
W = (G^{\dag}\otimes I) {\rm select}(V) (G\otimes I),
\end{aligned}
\end{equation}
which when applied to the input state $\ket{0}\otimes\ket{\Psi}$ gives
 \begin{equation}
 \begin{aligned}
W(\ket{0}\otimes \ket{\Psi}) = \frac{1}{s}(\ket{0}\otimes  \widetilde{U}(t)\ket{\Psi})+\frac{\sqrt{s^2-1}}{s} \ket{\phi},
\end{aligned}
\end{equation}
where $\ket{\phi}$ is orthogonal to the subspace $span\{\ket{0}\otimes \mathcal{H}_s \}$. After projecting to $\ket{0}$ of the ancillary system $\mathcal{H}_{an}$, we can obtain the approximate evolution $\widetilde{U}(t)\ket{\Psi}$ with probability $1/s$. However, the failure probability is nonignorable, thus we need to boost the coefficient before $\ket{0}\otimes  U(t)\ket{\Psi}$.
When $s=2$, we could use the oblivious amplitude amplification to boost the probability,
\begin{equation}
 \begin{aligned}
-WRW^{\dag}RW (\ket{0}\otimes \ket{\Psi})=\ket{0}\otimes  \widetilde{U}(t)\ket{\Psi},
\end{aligned}
\end{equation}
where $R$ is the reflection $R=(I-2\ket{0}\bra{0})\otimes I$ acting on $\mathcal{H}_{an} \otimes \mathcal{H}_s$.
Here oblivious amplitude amplification 
is similar to the iterations in the Grover algorithm. The operation $RW$ can be understood as the reflection along the axis vector $\ket{0}\otimes  U(t)\ket{\Psi}$. After this, we implement $-WRW^{\dag}=2 W\ket{0}\bra{0}W^{\dag} -I$, which is the reverse reflection along the axis $W(\ket{0}\otimes \ket{\Psi})$ and completes the amplification.
Thus the post selected evolution $\widetilde{V}(t)$ is
\begin{equation}
 \begin{aligned}
\widetilde{V}(t)= (\bra{0}\otimes I) -WRW^{\dag}RW (\ket{0}\otimes I).
\end{aligned}
\end{equation}

In general, $s$ can be far away from $2$. Similar to the product formula method, here we could divide the simulation time $t$ into $r$ segments such that  the simulation time  in each segment $\tau=t/r=\frac{\ln2}{\sum_{l}\alpha_l}$ satisfies
\begin{equation}
 \begin{aligned}
s=\sum_{k=0}^K\frac{(t\sum_{l}a_l/r)^k}{k!}\approx exp\left(t\sum_{l}a_l/r\right)=2.
\end{aligned}
\end{equation}
For the remaining time $\tau_{re}$ less than $\tau$, 
$\tau=\frac{\ln2}{\sum_{l}\alpha_l},
~t=(r-1) \tau+ \tau_{re}. $
Another ancillary qubit can be used to exactly boost the value of $s$ to $2$, see Refs~\cite{Childs9456,TaylorSeries}. 
Thus the whole evolution can be implemented by
\begin{equation}
 \begin{aligned}
\widetilde{U}_{TS}(t) = \widetilde{V}(\tau_{re})\widetilde{V}(\tau)^{r-1}.
\end{aligned}
\end{equation}
The algorithmic error mainly comes from the finite truncated Taylor series
\begin{equation}
\begin{aligned}
\|\widetilde{U}(\tau)-U_0(\tau)\|\le 2\frac{(\ln 2)^{K+1}}{(K+1)!}.
\end{aligned}
\end{equation}
It can be verify that
\begin{equation}
\begin{aligned}
\widetilde{V}(\tau)=\frac{3}{s}\widetilde{U}(\tau)-\frac{4}{s^3}\widetilde{U}(\tau)\widetilde{U}(\tau)^\dag \widetilde{U}(\tau).\\
\end{aligned}
\end{equation}
Supposing the truncated error $\delta=2\frac{(\ln 2)^{K+1}}{(K+1)!}$, the distance for each segment can be bound as
\begin{equation}
\begin{aligned}\label{Eq:segment}
\|\widetilde{V}(\tau)-U_0(\tau)\|\le \frac{\delta^2+3\delta+4}{2}\delta,
\end{aligned}
\end{equation}
and the total error accumulates with $r$ segments as
\begin{equation}
\begin{aligned}
\left\|\widetilde{V}(\tau_{re})\widetilde{V}(\tau)^{r-1}-e^{-iHt}\right\|\le \frac{\delta^2+3\delta+4}{2}\delta r.
\end{aligned}
\end{equation}
Given a fixed total allowed error $\varepsilon$, the error for each segment is $O(\varepsilon/r)$. In order to make Eq.~\eqref{Eq:segment} satisfy the error constraint, the parameter $K$ can be chosen as $K=O\left(\frac{\log \frac{\alpha t}{\varepsilon}}{\log \log \frac{\alpha t}{\varepsilon}}\right)$,
where $\alpha=\sum_{l}\alpha_l$.
The segment number is $r=\frac{\alpha t}{\ln2}$. In each segment, we need to implement $G$ and ${\rm select}(V)$.
The oracle $G$ can be implemented by a unary encoding which needs $O(KL)$ gates. And ${\rm select}(V)$ mainly contributes to the gate complexity, which can be implemented via 
$K$ controlled-${\rm select}(H)$ requiring the gate cost
$O(KL(n+\log L))$. Thus the total gate complexity is
$O\left( \alpha t L(n+\log L) \frac{\log \frac{\alpha t}{\varepsilon}}{\log \log \frac{\alpha t}{\varepsilon}}\right)$.
We refer to Refs~\cite{TaylorSeries,Childs9456} for more details of this algorithm.
\section{Reducing the errors in the original LCU scheme}\label{Appendix:reduce}
Here we give the refined error estimation for the Taylor series method using the second order, third order cancelling method. The fourth order cancelling is similar.
In the second order cancellation, we can divide the high order terms into two parts: odd order terms and even order terms. And the spectral norm of each part can be bound via Eq.~\eqref{Eq:Hmupp},
\begin{equation}
\begin{split}
&\left\|\widetilde{U}(t)-\sum_{k=0}^{\infty}\frac{(-itH)^k}{k!}\right\|\\
&\le \sum_{j=0,1} \left\|\sum_{k=K+1,~k=j~(mod ~2)}^{\infty}\frac{(-itH)^k}{k!}\right\| \\
&\le \sum_{j=0,1}\left(\sum_{k=K+1,~k=j~(mod ~2)}^{\infty}\frac{t^k \alpha_{comm}^{\lfloor \frac{k}{2}\rfloor}\alpha^j}{k!} \right)\\
&\le \frac{(t\sqrt{\alpha_{comm}})^{K+1}}{(K+1)!}\sum_{j=0,1}\bigg( \sum_{l\ge 0,~K+1+l=j~(mod ~2)}q^j\frac{t^{l}}{l!}\sqrt{\alpha_{comm}}^{l}\bigg)\\
&=\frac{(t\sqrt{\alpha_{comm}})^{K+1}}{2(K+1)!}\Bigg[q\left(e^{t\sqrt{\alpha_{comm}}}+(-1)^K e^{-t\sqrt{\alpha_{comm}}}\right)+\left(e^{t\sqrt{\alpha_{comm}}}+(-1)^{K-1} e^{-t\sqrt{\alpha_{comm}}}\right)  \Bigg]\\
&=\frac{(t\alpha)^{K+1}}{(K+1)!}\frac{(q+1) e^{t\alpha/q}+(-1)^K(q-1)e^{-t\alpha/q}  }{2q^{K+1}}.\\
\end{split}
\end{equation}
Here we give the explicit form of the error via third order cancellation. For simplicity, we suppose $K=2~(mod~ 3)$ and denote $\frac{\alpha}{(\alpha^{(3)})^{1/3}}=q_3$ and the trivial parameters $q_0=q_1=1$.
Then the error are divided into three parts as follows. The cases $K=0~or~1~(mod~ 3)$ are similar.
\begin{equation}
 \begin{split}\label{Eq:Newerror}
&\left\|\widetilde{U}(t)-\sum_{k=0}^{\infty}\frac{(-itH)^k}{k!}\right\|\\
&\le \sum_{j=0,1,2}\left\|\sum_{k=K+1,~k=j~(mod ~3)}^{\infty}\frac{(-itH)^k}{k!}\right\| \\
&\le \sum_{j=0,1,2} \sum_{k=K+1,~k=j~(mod ~3)}^{\infty}\frac{t^k (\alpha^{(3)})^{\lfloor \frac{k}{3}\rfloor}\alpha^{(j)}}{k!}\\
&\le \frac{[t(\alpha^{(3)})^{1/3}]^{K+1}}{(K+1)!}\sum_{j=0,1,2}\bigg( \sum_{l=j~(mod~3),~l\ge 0} (\frac{q_3}{q_j})^{j}\frac{t^{l}}{l!}(\alpha^{(3)})^{l/3}\bigg)\\
&=\frac{(t\alpha)^{K+1}}{(K+1)!}\frac{
S_0+q_3S_1+(\frac{q_3}{q_2})^2S_2}{q_3^{K+1}},\\
\end{split}
\end{equation}
where 
\begin{equation}
    \begin{aligned}
 S_0&=\sum_{l=0~(mod~3),~l\ge 0} \frac{t^{l}}{l!}(\alpha^{(3)})^{l/3}=\frac{1}{3}
 (e^{t\alpha/q_3}+e^{\omega t\alpha/q_3}+e^{\omega^2 t\alpha/q_3}),\\
  S_1&=\sum_{l=1~(mod~3),~l\ge 0} \frac{t^{l}}{l!}(\alpha^{(3)})^{l/3}=\frac{1}{3}
 (e^{t\alpha/q_3}+\omega^2 e^{\omega t\alpha/q_3}+\omega e^{\omega^2 t\alpha/q_3}),\\
   S_2&=\sum_{l=2~(mod~3),~l\ge 0} \frac{t^{l}}{l!}(\alpha^{(3)})^{l/3}=\frac{1}{3}
 (e^{t\alpha/q_3}+\omega e^{\omega t\alpha/q_3}+\omega^2 e^{\omega^2 t\alpha/q_3}),\\
  &\omega^2+\omega+1=0.\\ 
    \end{aligned}
\end{equation}
For a $p$-th order cancellation, we denote the parameter  $\frac{\alpha}{(\alpha^{(p)})^{1/p}}=q_p$ and could give a general result,
\begin{equation}\label{Eq:p-Newerror}
 \begin{split}
&\left\|\widetilde{U}(t)-\sum_{k=0}^{\infty}\frac{(-itH)^k}{k!}\right\|\\
&\le \sum_{j=0}^{p-1}\left\|\sum_{k=K+1,~k=j~(mod ~p)}^{\infty}\frac{(-itH)^k}{k!}\right\| \\
&\le \sum_{j=0}^{p-1} \sum_{k=K+1,~k=j~(mod ~p)}^{\infty}\frac{t^k (\alpha^{(p)})^{\lfloor \frac{k}{p}\rfloor}\alpha^{(j)}}{k!}\\
&\le \sum_{j=0}^{p-1} \sum_{k=K+1,~k=j~(mod ~p)}^{\infty}\frac{\alpha^{(j)}}{(\alpha^{(p)})^{j/p}}\frac{t^k (\alpha^{(p)})^{ \frac{k}{p}}}{k!}\\
&\le \frac{[t(\alpha^{(p)})^{1/p}]^{K+1}}{(K+1)!} \max_j(\frac{q_p}{q_j})^j
 \sum_{l\ge 0}\frac{t^{l}}{l!}(\alpha^{(p)})^{l/p}\\
&\le \frac{(t\alpha)^{K+1}}{(K+1)!}\frac{
e^{t\alpha/q_p}}{q^{K+1}_p}\max_j(\frac{q_p}{q_j})^j.\\
\end{split}
\end{equation}

\section{Details of modified LCU Algorithm reducing high order error}\label{App:modif}
In the modified LCU method, we construct an approximate $K$-th order term as
\begin{equation}
\begin{aligned}
\widetilde{R}_{K}
&=\frac{(-itH)^{K-1}}{K!}\left[ \sum_{l=1}^L{\gamma_l}(-i H_l) + \sum_{j=0}^{E} \widetilde{\gamma_{j}} U_{j} \right].\\
\end{aligned}
\end{equation}
According to the anti-commutative cancellation, the second and third order terms $H^2$ and $H^3$ in $\sum_{\vec{e}} \widetilde{\gamma_{\vec{e}}}  H_{\vec{e}}$ are
\begin{equation}
 \begin{aligned}
H^2&=  \beta_0 I+\sum_{(l_1,l_2)\in Comm^* } \alpha_{l_1}\alpha_{l_2}H_{l_1}H_{l_2},\\
H^3&=\sum_l \beta_l H_l + \sum_{(l_1,l_2),(l_1,l_3),(l_2,l_3)\in Comm^*} \alpha_{l_1}\alpha_{l_2}\alpha_{l_3}H_{l_1}H_{l_2}H_{l_3}
\end{aligned}
\end{equation}
where $Comm^*=\{(i,j)|[H_i,H_j]=0,~ i\neq j\}$.
Here the first line $H^2$ is directly from the anti-commutative cancellation. 
For $H^3$ terms, when $l_1,l_2,l_3$ are  equal $l_1=l_2=l_3$ or two of them are equal, for example, $l_1=l_2\neq l_3$. Then we have
$H_{l_1}H_{l_2}H_{l_3}=H_{l_3}$.
When $l_1,l_2,l_3$ are different,
there are two cases: 1. All the pairs are commutative, 2. There is at least one pair $H_{l_1}$ and $H_{l_2}$ which are anti-commutative. 
For case 2, all the possible six sequences can be cancelled pair by pair as 
\begin{equation}
    \begin{aligned}
H_{l_1}H_{l_2}H_{l_3}+ H_{l_2}H_{l_1}H_{l_3}=0,\\
H_{l_1}H_{l_3}H_{l_2}+ H_{l_2}H_{l_3}H_{l_1}=0,\\
H_{l_3}H_{l_2}H_{l_1}+ H_{l_3}H_{l_1}H_{l_2}=0.
    \end{aligned}
\end{equation}
Then only the terms  
$H_{l_1}H_{l_2}H_{l_3}$ with
$(l_1,l_2),(l_1,l_3),(l_2,l_3)\in Comm^*$ (case 1) are remained. 

The upper bound of $(K+1)$-th order terms is straightforward according to the definition of $E_{\epsilon}$, which are the terms not included in $U_j$ and $-iH_l$.
For the $K+2$-th order terms,
$\beta_l H_l$ are already included into $\gamma_l (-iH_l)$ and the remaining terms are
\begin{equation}
 \begin{aligned}
&E_{K+2}=\frac{t^{K+2}}{(K+2)!} (-iH)^{K-1}
\sum_{(l_1,l_2),(l_1,l_3),(l_2,l_3)\in Comm^*}\alpha_{l_1}\alpha_{l_2}\alpha_{l_3}(-i)^3H_{l_1}H_{l_2}H_{l_3}.
\end{aligned}
\end{equation}
We denote the coefficient
\begin{equation}
 \begin{aligned}
\alpha^{(3)}_r= \sum_{(l_1,l_2),(l_1,l_3),(l_2,l_3)\in Comm^*} \alpha_{l_1}\alpha_{l_2}\alpha_{l_3}iH_{l_1}H_{l_2}H_{l_3},
\end{aligned}
\end{equation}
and we have
\begin{equation}
 \begin{aligned}
\left\|E_{K+2}\right\|\le\frac{t^{K+2}}{(K+2)!} \|H^{K-1}\| \alpha^{(3)}_r.
\end{aligned}
\end{equation}
When $K-1$ is even, we use the bound 
\begin{equation}
   \|H^{K-1}\|\le \alpha_{comm}^{ \frac{K-1}{2}} 
\end{equation}
in Eq.~\eqref{Eq:Hmupp}. 
We could also use the similar method to carefully calculate the rest high order part 
$\|E_{\infty}\|$, but it would not influence the error a lot. Thus we directly use the bounds in Eq.~\eqref{Eq:Hmupp} and obtain 
\begin{equation}
    \begin{aligned}
&\|E_{\infty}\|\le \frac{t^{K+3}\sqrt{\alpha_{comm}}^{K+3}}{(K+3)!}
\Bigg[\frac{\alpha}{\sqrt{\alpha_{comm}}}\bigg(\sum_{l~odd,~l\ge 0} \frac{t^{l}}{l!}\sqrt{\alpha_{comm}}^{l}\bigg)
+ \sum_{l~even,~l\ge 0} \frac{t^{l}}{l!}\sqrt{\alpha_{comm}}^{l}\Bigg].
\end{aligned}
\end{equation}
In the modified scheme,  all the low order $K-1$ order terms can be  implemented as usual.
To implement modified $\widetilde{R}_{K}$, we need to replace the $K$-th controlled ${\rm select}(H)$ with one controlled ${\rm select}(H')$, where ${\rm select}(H')$ has $E+1+L$ unitaries to select, which is more complicate than ${\rm select}(H)$,
 \begin{equation}
\begin{aligned}
&{\rm select}(H)= \sum_{l=1}^{L}\ket{l}\bra{l}\otimes(-iH_l),\\
&{\rm select}(H')= \sum_{l=1}^{L}\ket{l}\bra{l}\otimes(-iH_l)+
 \ket{L+1}\bra{L+1} \otimes(-I)\\
 &+ \sum_{j=1}^{E}\ket{L+1+j}\bra{L+1+j} \otimes U_{j}.\\
\end{aligned}
\end{equation}

In Ref.~\cite{Childs9456}, a ${\rm select}(H)$ with $L$ different unitaries to select  requires
$7.5\cdot 2^w+6w-26$ CNOT gates and $7.5\cdot 2^w+6w-28$ T gates, where
$w= \lceil \log_2 L\rceil$.
Then we can see that when $E\le 2^w-L-1$, ${\rm select}(H)$ and 
${\rm select}(H')$ have the same gate cost. Thus, in the numerical simulation, we implement the modified schemes with $E=0, 2^w-L-1$, which indicates almost no additional gate cost.

\section{Modified LCU method for the pair-wisely anti-commutative Hamiltonians}\label{App:mutually}
\subsection{Proof of Lemma \ref{Le:perfect}}
We can prove it iteratively. The conclusion is true for $m=1$. Now, suppose it is true for $m>1$, we have
\begin{equation}
    \begin{aligned}
    &H^{m+1}=H^{m}H\\
    &=\bigg(\gamma^{(m)}_0 I +\sum_{l=1}^{L}\gamma^{(m)}_l H_{l}\bigg)\bigg(\sum_{l=1}^{L}\alpha_l H_{l}\bigg)\\
    &=\sum_{l=1}^{L} \gamma^{(m)}_l\alpha_l I + \gamma^{(m)}_0\sum_{l=1}^{L}\alpha_l H_{l} + \sum_{l_1\neq l_2}\gamma^{(m)}_{l_1}\alpha_{l_2} H_{l_1}H_{l_2}\\
    \end{aligned}
\end{equation}
The third term is 0 due to anti-commutative relation and we have
\begin{equation}
    \gamma^{(m+1)}_0 = \sum_l \gamma^{(m)}_l\alpha_l, \, \gamma^{(m+1)}_l = \gamma^{(m)}_0\alpha_l.
\end{equation}
Thus $H^{m+1}= \gamma^{(m+1)}_0 I+\sum_{l=1}^{L}\gamma^{(m)}_l H_{l} $.

\subsection{Implementation of LCU}\label{Sec:LCUimple}
From Lemma~\ref{Le:perfect}, we can expand the evolution $U(t)=e^{-iHt}$ as
\begin{equation}
\begin{aligned}
U(t)=\widetilde{U}(t)&= \tilde{\alpha}_0 I+ \sum_{l} \tilde{\alpha}_l (-iH_l),\\
\end{aligned}
\end{equation}
where  
\begin{equation}
\begin{aligned}
\tilde{\alpha}_0&=1-\beta_s^2t^2/2!+\beta_s^4t^4 /4!+\dots=\cos (t\beta_s),\\
\tilde{\alpha}_l&=\alpha_lt-\alpha_l\beta_s^2t^3/3!+ \alpha_l\beta_s^4 t^5/5!+\dots\\
&=\frac{\alpha_l}{\beta_s} \sin(t\beta_s), 
~l=1,\dots,L.\\
\beta_s&=\sqrt{\sum_{l=1}^{L}\alpha_l^2}.
\end{aligned}
\end{equation}
Now we show how to implement the evolution via the LCU method. We show two different methods here.
The first method is via direct implementation of $U(t)$ via LCU. 
Similar to the original LCU method, we need oracles $G$, ${\rm select}(U)$, \begin{equation}
 \begin{aligned}
&G\ket{0}=\frac{1}{\sqrt{s}}\sum_{j}\sqrt{|\tilde{\alpha}_l|}\ket{j},\\
& {\rm select}(U)=\sum_l \ket{l}\bra{l}\otimes U_l,\\
 &W=(G^{\dagger}\otimes I){\rm select}(U)(G\otimes I),\\
    &s=\sum_{l}|\tilde{\alpha}_l|=|\cos (t\beta_s)|+\frac{\alpha}{\beta_s}|\sin (t\beta_s)|.
\end{aligned}
\end{equation}
Here we choose $U_0=I$ or $-I$, $U_l=-iH_l$ or $iH_l$ according to the sign of $\tilde{\alpha}_l$. 
According to Ref.~\cite{TaylorSeries}, ${\rm select}(U)$ can be implemented via ${\rm select}(U_l)$ one by one (totally $L+1$). And each  ${\rm select}(U_l)$ requires $O(n+\log L)$ gates where $n$ is the number of qubits in the target system. 
Thus we need $ O(L(n+\log L))$ gates to implement $W$.
Similar to the original method, we also need multiple use of $W$ and reflection to boost the coefficient before $\ket{0}\otimes  U(t)\ket{\Psi}$.
According to Ref.~\cite{berry2014exponential}, we require $ O(s)$ number of repetition use of $W$. 
Due to $1\le s\le 1+\sqrt{L}$, the total gate complexity is
$ O(L^{\frac{3}{2}}(n+\log L))$, which is only related to the number of terms used in LCU $L$, not related to $t$ and $\varepsilon$.

The second method uses a few different segments to simulate the whole evolution. 
From Eq.~\eqref{Eq:perefects}, when $s=1$ we do not need to use the oblivious amplitude amplification, $U(t)$ can be implemented by
 \begin{equation}
 \begin{aligned}
W(\ket{0}\otimes \ket{\Psi}) = \ket{0}\otimes  U(t)\ket{\Psi}.
\end{aligned}
\end{equation}
Thus we choose the first segment as $t_1=\lfloor\frac{t\beta_s}{\pi}\rfloor\frac{\pi}{\beta_s}$ such that $s=1$. The rest evolution time is 
$t_2=t-t_1\le \frac{\pi}{\beta_s}$ and the corresponding $s$ can still be very large.
When $\frac{\alpha^2}{\beta_s^2}\ge 3$, we choose another simulation time $t_{seg}$ as 
\begin{equation}
\begin{aligned}
t_{seg}=\frac{\arcsin\left[\frac{2\frac{\alpha}{\beta_s}-\sqrt{\frac{\alpha^2}{\beta_s^2}-3}}{1+\frac{\alpha^2}{\beta_s^2}}\right]}{\beta_s}
\end{aligned}    
\end{equation}
such that  
\begin{equation}
\begin{aligned}
    s&=|\cos (t_{seg}\beta_s)|+\frac{\alpha}{\beta_s}|\sin (t_{seg}\beta_s)|=2.
\end{aligned}    
\end{equation}
And $t_2$ is divided into $r$ segments $r=\lfloor\frac{t_2}{t_{seg}}\rfloor$, $t_2=rt_{seg}+t_{rest},t_{rest}\le t_{seg}$. 
In the rest time, we have
\begin{equation}
\begin{aligned}
    1<s=|\cos (t_{rest}\beta_s)|+\frac{\alpha}{\beta_s}|\sin (t_{rest}\beta_s)|<2.
\end{aligned}    
\end{equation}
Thus we need $r+2$ segments, where $r$ can be bound as 
\begin{equation}
r\le \frac{\pi}{\arcsin\left[\frac{2\frac{\alpha}{\beta_s}-\sqrt{\frac{\alpha^2}{\beta_s^2}-3}}{1+\frac{\alpha^2}{\beta_s^2}}\right]}=\mathcal O(\sqrt{L})   
\end{equation}
The last equation is due to $ \frac{\alpha^2}{\beta_s^2}\le L$. 

When $\frac{\alpha^2}{\beta_s^2}< 3$, we also have $1<s<2$. For all the segments where $s<2$,  
we can use another aniclla to convert $s<2$ into the $s=2$ case~\cite{TaylorSeries,Childs9456}. And all the $s=2$ segments can be implemented via the oblivious amplitude amplification as the usual LCU method.  
Consequently, the upper bound of gate cost is 
$r \mathcal O(L(n+\log L))=\mathcal O(L^{\frac{3}{2}}(n+\log L))$.
We could implement this digital simulation efficiently without algorithmic error via standard LCU method.


\subsection{Proof of Corollary \ref{Cor:nearp}}
Here we consider the case where the Hamiltonian $H$ is close to the perfectly anti-commutative,
\begin{equation}
\begin{aligned}
    H^2=\beta_s^2 I +V, ~~\|V \|\le \varepsilon_{A}.
\end{aligned}
\end{equation}
Then the unitary evolution can be expressed as
\begin{equation}
\begin{aligned}
U_0(t)&=\sum_{k=0}^{\infty}\frac{(-itH)^k}{k!}\\
&= \sum_{even~k\ge 0} \frac{t^k}{k!}(-\beta_s^2I-V)^{\frac{k}{2}}+ \sum_{odd~k\ge 1} \frac{t^k}{k!}(-\beta_s^2I-V)^{\lfloor\frac{k}{2}\rfloor}\sum_{l}(-iH_l).\\
\end{aligned}
\end{equation}
Note that here $V$ is not unitary but the sum of a few different unitaries.
Here we also implement the operator $\widetilde{U}(t)$ in the perfect case in Eq.~\eqref{Eq:perfectU}
\begin{equation}
\begin{aligned}
&\widetilde{U}(t)= \tilde{\alpha}_0 I+ \sum_{l} \tilde{\alpha}_l (-iH_l),\\
\end{aligned}
\end{equation}
and the error can be bound by
\begin{equation}
\begin{aligned}
&\left\|\widetilde{U}(t)-\sum_{k=0}^{\infty}\frac{(-itH)^k}{k!}\right\|\\
&\le \sum_{even~k\ge 2} \frac{t^k}{k!}[(\beta_s^2+\varepsilon_A)^{\frac{k}{2}}-\beta_s^k]+ \sum_{odd~k\ge 3} \frac{t^k}{k!}\alpha[(\beta_s^2+\varepsilon_A)^{\lfloor\frac{k}{2}\rfloor}-\beta_s^{k-1}]\\
&\le \frac{1}{2}\left[ \left(e^{t\sqrt{\beta_s^2+\varepsilon_A}}+ e^{-t\sqrt{\beta_s^2+\varepsilon_A}}\right)-   
\left(e^{t\beta_s}+e^{-t\beta_s}\right)\right]\\
&+\frac{1}{2}\bigg[\frac{\alpha}{\sqrt{\beta_s^2+\varepsilon_A}}\left(e^{t\sqrt{\beta_s^2+\varepsilon_A}}- e^{-t\sqrt{\beta_s^2+\varepsilon_A}}\right)-\frac{\alpha}{\beta_s}\left(e^{t\beta_s}-e^{-t\beta_s}\right) \bigg],
\end{aligned}
\end{equation}
which is roughly $\mathcal O(t^2\varepsilon_A)+\mathcal O(t^3\varepsilon_A\alpha)$ with a small $t$. 
The first inequality is because high order of $\beta_s^2I$ are all included in the coefficient $\gamma_0$.


\end{appendix}

\bibliographystyle{plainnat}
\bibliography{bibTaylor}


\end{document}